\documentclass[sigconf,nonacm]{acmart}
\renewcommand\footnotetextcopyrightpermission[1]{} 
\setcopyright{none}

\AtBeginDocument{%
  }

\newtheorem{hyp}{Hypothesis}

\setcopyright{none}
\usepackage{subcaption}
\usepackage{array}
\usepackage{multirow}
\usepackage{comment}
\usepackage[table]{xcolor}
\usepackage[most]{tcolorbox}
\usepackage{pifont} 
\usepackage{enumitem} 
\usepackage{threeparttable}
\usepackage{tabularx}
\usepackage{ragged2e}

\newenvironment{takeaway}
  {\begin{tcolorbox}[colback=gray!5, colframe=blue!40]}
  {\end{tcolorbox}}

\begin{document}

\title{SMART: A Social Movement Analysis \& Reasoning Tool \\ with Case Studies on \#MeToo and \#BlackLivesMatter}

\author{Valerio La Gatta}
\authornote{Both authors contributed equally to this research.}
\orcid{0000-0003-1470-8053}
\affiliation{%
  \institution{Northwestern University}
  \department{Department of Computer Science and Buffett Institute for Global Affairs}
  \city{Evanston}
  \state{Illinois}
  \country{USA}
}
\email{valerio.lagatta@northwestern.edu}

\author{Marco Postiglione}
\authornotemark[1]
\orcid{0000-0003-1470-8053}
\affiliation{%
  \institution{Northwestern University}
  \department{Department of Computer Science and Buffett Institute for Global Affairs}
  \city{Evanston}
  \state{Illinois}
  \country{USA}
}
\email{marco.postiglione@northwestern.edu}

\author{Jeremy Gilbert}
\orcid{0009-0008-2585-0972}
\affiliation{%
  \institution{Northwestern University}
  \department{Medill School of Journalism, Media, Integrated Marketing Communications}
  \city{Evanston}
  \state{Illinois}
  \country{USA}
}
\email{jeremy.gilbert@northwestern.edu}

\author{Daniel W. Linna Jr.}
\orcid{0000-0003-0053-5198}
\affiliation{%
  \institution{Northwestern University}
  \department{Pritzker School of Law and Department of Computer Science}
  \city{Evanston}
  \state{Illinois}
  \country{USA}
}
\email{dan.linna@northwestern.edu}

\author{Morgan Manella Greenfield}
\affiliation{%
  \institution{The Wall Street Journal}
  \city{New York}
  \country{USA}
}
\email{morgan.greenfield@wsj.com}

\author{Aaron Shaw}
\orcid{0000-0003-4330-957X}
\affiliation{
  \institution{Northwestern University}
  \department{Department of Communication Studies}
  \city{Evanston}
  \state{Illinois}
  \country{USA}
}
\email{aaron.shaw@northwestern.edu}

\author{V.S. Subrahmanian}
\orcid{0000-0001-7191-0296}
\affiliation{%
  \institution{Northwestern University}
  \department{Department of Computer Science and Buffett Institute for Global Affairs}
  \city{Evanston}
  \state{Illinois}
  \country{USA}
}
\email{vss@northwestern.edu}

\renewcommand{\shortauthors}{Valerio La Gatta et al.}

\begin{abstract}
  Social movements supporting the UN's Sustainable Development Goals (SDGs) play a vital role in improving human lives. If journalists were aware of the relationship between social movements and external events, they could provide more precise, time-sensitive reporting about movement issues and SDGs. Our SMART system achieves this goal by collecting data from multiple sources, extracting emotions on various themes, and then using a transformer-based forecasting engine (DEEP) to predict quantity and intensity of emotions in future posts.  This paper demonstrates SMART's \emph{Retrospective} capabilities required by journalists via case studies analyzing social media discussions of the \#MeToo and \#BlackLivesMatter before and after the 2024 U.S. election. We create a novel 1-year dataset which we will release upon publication. It contains over 2.7M Reddit posts and over 1M news articles.  We show that SMART enables early detection of discourse shifts around key political events, providing journalists with actionable insights to inform editorial planning. SMART was developed through multiple interactions with a panel of over 20 journalists from a variety of news organizations over a 2-year period, including an author of this paper.
\end{abstract}



\maketitle


\section{Introduction}

Social movements (SMs) play a vital role in advancing the United Nations' Sustainable Development Goals (SDGs) by mobilizing public attention, challenging institutional practices, and catalyzing policy change \cite{nathanson1999social}. The \#MeToo SM influenced reporting patterns of gender-related crimes \cite{canaveras2024role} and brought attention to health effects linked to sexual harassment \cite{o2018metoo}. The Black Lives Matter (BLM) SM reshaped policing practices, criminal justice reforms, and public discourse on racial inequality across multiple nations \cite{brown2025happened}.

Journalists serve as intermediaries between SMs and the public, translating grassroots mobilization into narratives that shape public understanding and institutional responses. However, journalists currently lack systematic tools to understand the linkage between SMs and major events (e.g., political developments), or anticipate emerging discourse shifts before they become widely visible. This gap has significant consequences in an era of rapidly evolving online discourse: \textit{by the time a surge in SM activity becomes obvious, the opportunity for early, informed journalistic coverage may have passed. Journalists want to cover important social phenomena as early as possible}.

To address this need, we developed SMART (Social Movement Analysis and Reasoning Tool), a system for tracking SDG-supportive social movements across multiple data sources. SMART emerged from a two-year collaboration with approximately 20 journalists from a variety of news organizations, including The Wall Street Journal, the American Press Institute, and The Washington Post. One of these journalists is a co-author of this work and articulated four core requirements (see the Appendix for the full list of collaborators): \textsf{(i)} identifying SMs rising in popularity measured by discourse volume, \textsf{(ii)} understanding how sentiments and emotions evolve within SM discourse over time, \textsf{(iii)} examining relationships between key political events and SM dynamics, and \textsf{(iv)} detecting discourse shifts early enough to enable proactive editorial planning. While our prior work introduced our DEEP module \cite{697729bb748d292568f980b9cc6677a7ddc618a9} to address \textsf{(iv)}, the forecasting dimension, this paper focuses on SMART's retrospective analysis capabilities, \textsf{(i)}--\textsf{(iii)} above that enable journalists to investigate historical patterns in SM discourse.

We demonstrate SMART's capabilities through case studies analyzing \#MeToo and Black Lives Matter (BLM) discourse before and after the 2024 U.S. presidential election. These SMs represent distinct but complementary dimensions of social justice: \#MeToo centers on gender-based violence and institutional accountability (SDG \#5 on Gender Equality), while BLM focuses on racial justice and police reform (SDG \#10 on Reducing Inequality). The 2024 US election cycle provides a natural setting to examine how political events structure SM discourse, as both SMs intersected with campaign rhetoric, candidate positions, and post-election policies.

Our rigorous statistical analysis tests five hypotheses about the relationship between key political events (KPEs) and SM discourse for \#MeToo and BLM\footnote{Like most data-driven work, we study statistical linkage, not causation.}. We encourage readers to consider their expectations for each of the following questions:
\begin{itemize}
\item[\textbf{H1}] Does the volume of SM discourse increase significantly during the days surrounding KPEs compared to other periods?

\item[\textbf{H2}] Is the volume of SM discourse higher in the days before KPEs than in the days after, indicating anticipatory, not reactive, patterns?

\item[\textbf{H3}] Do individual KPEs vary in their impact, with some increasing SM discourse volume and others decreasing it?

\item[\textbf{H4}] Does emotion intensity in SM discourse increase during the days around KPEs compared to baseline periods?

\item[\textbf{H5}] Is emotion intensity in SM discourse higher in the days before KPEs than in the days after? 

\end{itemize}

\section{Related Work}

Early systems like TwitInfo \cite{1cf2f8804eaa3050b6e8da58ed241eb490143f27} pioneered social media monitoring by visualizing Twitter events through peak detection and sentiment analysis. Subsequent work expanded into event detection \cite{ece6a2d9d932170e0825ce215a27e0ec06eaf2bb,722543bda19bfa306fbc8a581681cbada8b45a23}, security monitoring \cite{1a501dc3a9ca16b5004d26a11ab0626c1c51ab16}, and interactive visual analytics \cite{e30a6f7cf79517e84f796600cb17e0811b0c0558,4d1dec2e48af258ded1985c5e9762ccab77ca6fb}. Emergency response systems like CitizenHelper-Adaptive \cite{85021c6bae5f233104e6ef8259e1e356db4acf48} and Crowd4SDG \cite{dc29eff0af56107e52c3fcc0b5f015f0afe2199e} demonstrated the value of combining classifiers with human-in-the-loop adaptation during crises, while Hamad et al. \cite{eef3a2f908bdb6bea8355cb2ca0db2895a0da5ca} integrated multiple data streams for real-time misinformation tracking during riots.

Computational analyses of \#BlackLivesMatter and \#MeToo have established that online discourse is tightly coupled to offline events. Research has examined participation patterns and their links to protests \cite{9e6d7ed67c13841a3aad2674101cf454517c16a7}, event-driven recruitment dynamics \cite{a7c22d6e2dfe5dd46bb7d9bed1bb97cebb200ce9}, activist narrative diffusion \cite{9b19a1569504cc17bfe1675bb578f5b447fe6025}, and counter-movement framing \cite{9de092a71dc132c0428a9771fa687c4cd4160ddb,c4ccdf58ea07367fd228539801e286b140026741,25cab71ca4f3da15bb3f2485afe8c78020a2b32e}. Other studies model reciprocal relationships between online support and offline confrontations \cite{0e51a7d5a4840cdcac554c5dfeb35d14024de40a}, cross-national protest signatures \cite{c6911c929c1de0546e08ce511407af85a0db4b90}, and temporal trajectories of different discourses around key revelations \cite{ccbb2d947ed1688440ba459b3fcba4d3e593f98b}. While these studies provide valuable insights into individual movements, they typically focus on single platforms and lack systematic cross-platform comparisons of how discourse evolves differently across media types.

Furthermore, most social movement analyses rely on completely automated  hashtag- and keyword-based methods to collect relevant data \cite{9e6d7ed67c13841a3aad2674101cf454517c16a7,ccbb2d947ed1688440ba459b3fcba4d3e593f98b,a7c22d6e2dfe5dd46bb7d9bed1bb97cebb200ce9,9de092a71dc132c0428a9771fa687c4cd4160ddb,c4ccdf58ea07367fd228539801e286b140026741,c6911c929c1de0546e08ce511407af85a0db4b90,9b19a1569504cc17bfe1675bb578f5b447fe6025}. However, Bozarth et al. \cite{3a97dcd56b69eed4696648cf89943167be807473} demonstrate that unrepresentative keyword sets can lead to flawed conclusions, and Srikanth et al. \cite{8bccf634a1449901bba8abe437e9230f586b22b5} highlight the importance of human-assisted methodologies. \newline 

\noindent \textbf{Our Contributions. \;} SMART addresses these gaps with contributions aligned with social impact: \textsf{(I)} A novel dataset spanning one year (September 2024 - August 2025) containing over 2.7M Reddit posts and 1M news articles related to \#MeToo and BLM, which we will release upon publication. \textsf{(II)} A co-design methodology demonstrating how sustained collaboration with journalists (over 20 journalists across multiple organizations) shaped system requirements and evaluation criteria, providing a model for building socially-impactful research tools. \textsf{(III)} The first systematic cross-platform comparison of temporal dynamics of Reddit versus News media  for SMs, revealing fundamental differences in how these platforms are linked to \#MeToo and BLM discourse. \textsf{(IV)} An event impact framework operationalizing "key political events" (KPEs) for retrospective SM analysis, revealing substantial heterogeneity in how individual events affect discourse (some triggering increases, others decreases, and many showing no significant effect).

\section{SMART: An Overview}

\begin{figure}
    \centering
    \includegraphics[width=\linewidth]{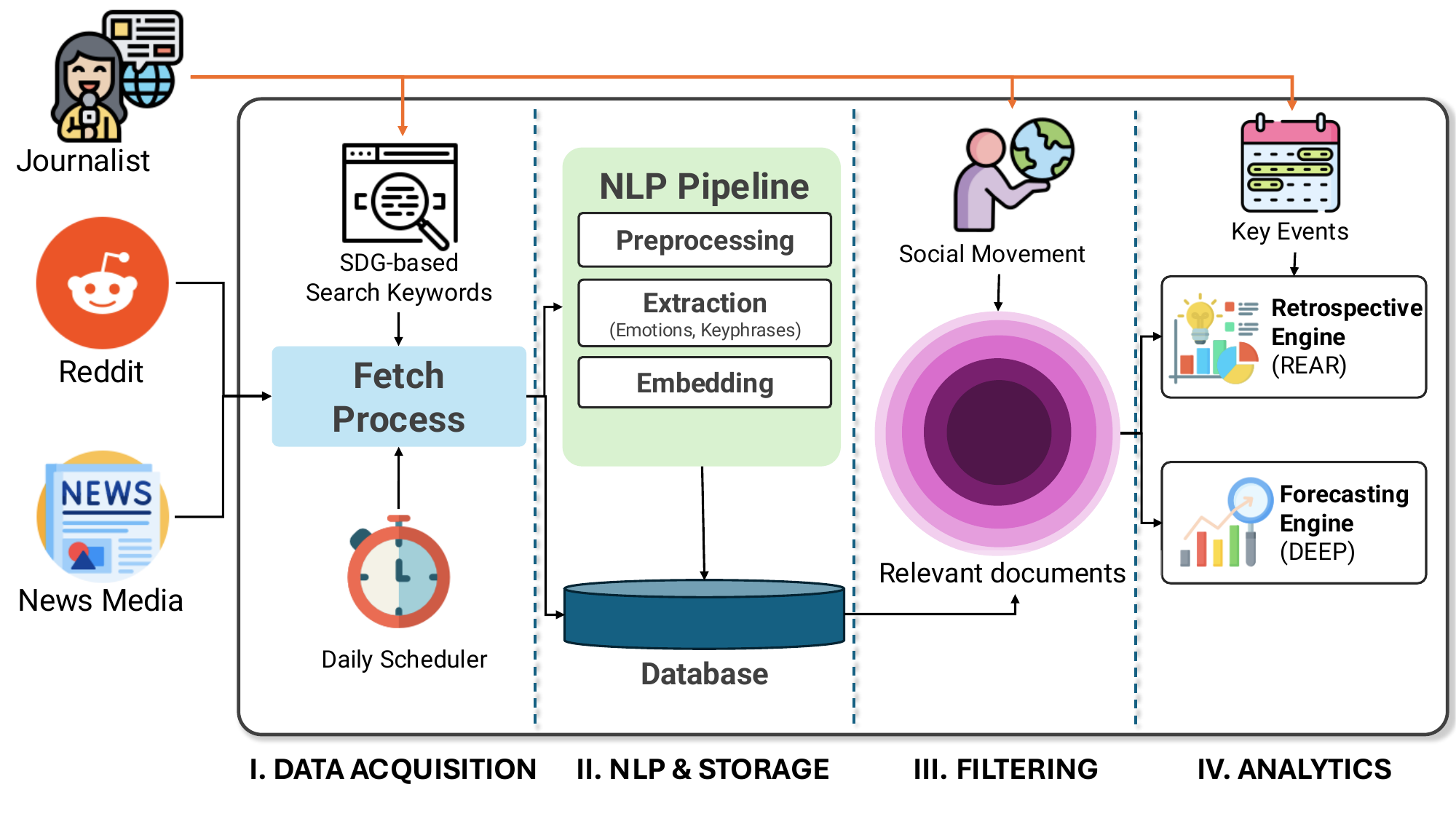}
    \caption{\textbf{SMART System Architecture Overview.}} 
    \label{fig:placeholder}
\end{figure}

SMART uses a multi-stage pipeline designed to support journalists investigating SMs via 4 components (Figure~\ref{fig:placeholder} shows the system architecture).  \textsf{(i)} Daily data acquisition from Reddit and news sources using SDG-specific keywords, \textsf{(ii)} NLP processing and multi-database storage, \textsf{(iii)} Multi-layered filtering to isolate SM-related discourse, and \textsf{(iv)} Dual analytics engines for retrospective analysis and forecasting. We now describe each component in detail.\footnote{This paper focuses on SMART as a whole and retrospective analysis.}

\subsubsection*{I. Data Acquisition}
SMART uses a keyword-driven data collection strategy operating on a daily schedule. A curated set of SDG-specific search terms, defined by journalists (who wanted this ability), drives queries across Reddit and News sources via the WorldNewsAPI\footnote{\url{www.worldnewsapi.com}}. Reddit posts are retrieved daily using PRAW\footnote{\url{https://github.com/praw-dev/praw}} with a 24-hour time filter, while news articles are fetched using date-restricted queries. For each target SDG, the system iterates through predefined keywords, collecting Reddit posts, news articles, and associated YouTube videos embedded in Reddit posts. 


\subsubsection*{II. NLP \& Storage}

Raw documents undergo multi-stage NLP processing before storage\footnote{{ Our pipeline ingests more than 50k documents per day, posing significant scalability constraints that make end-to-end LLM-based automation impractical.}}. Keyword extraction leverages KeyBERT \cite{grootendorst2020keybert} with the all-MiniLM-L6-v2 model \cite{reimers-2019-sentence-bert} and Amazon Comprehend\footnote{\url{https://aws.amazon.com/comprehend/}} to identify salient terms beyond the initial search keywords. Raw documents are stored in MongoDB with metadata (e.g., timestamps, source identifiers) stored in a linked relational database. Text embeddings are generated and stored in ChromaDB for semantic search capabilities, with batch processing (20,000 documents per batch) to optimize memory usage. Emotion analysis was computed with an open-source RoBERTa-base model\footnote{\url{https://huggingface.co/SamLowe/roberta-base-go_emotions}} finetuned on the GoEmotions dataset \cite{DBLP:conf/acl/DemszkyMKCNR20}. The system maintains separate collections for Reddit posts, news articles, and YouTube videos, with cross-references preserved through document IDs.

\subsubsection*{Filtering}

The filtering process starts when a journalist seeks to investigate a specific SM (e.g., \#MeToo). After data acquisition and NLP processing, documents undergo a multi-layered filtering process to construct datasets centered on the target SM while capturing varying degrees of semantic relevance. The first layer $L_0$ identifies documents that explicitly reference the target SM ($sm$). $L_0$ is the set of documents containing the SM keyword (e.g., ``\#MeToo'') either in the title or body. To capture documents that contribute to the movement's discourse without explicit mentions, we implement a keyword co-occurrence strategy for layers $L_1$ through $L_N$ using previously-extracted keywords (Step II, \textit{NLP \& Storage}). We first compute the frequency distribution of each word that co-occurs with $sm$. Keywords falling within the 99th percentile of co-occurrence form a high-salience vocabulary set that captures the core conceptual space of the movement. Documents are then assigned to subsequent layers based on what proportion of this high-salience vocabulary appears in each document. Specifically, layer $L_1$ contains documents where at least 40\% of the high-salience keywords are present, $L_2$ requires 35\%, $L_3$ requires 30\%, $L_4$ requires 25\%, $L_5$ requires 20\%, $L_6$ requires 15\%, $L_7$ requires 10\%, and $L_8$ requires 5\%.  This progressive threshold relaxation balances precision and recall, enabling inclusion of semantically related documents at varying distances from core terms used in that SM's discourse. For our empirical analyses in Section~\ref{sec:case_studies}, we use layer $L_5$ for both SMs and Platforms, as this layer provides a balance between semantic relevance and sufficient sample size across all datasets. {See Appendix \ref{app:ml_stra} for further hyper-parameters details. }


\subsubsection*{Analytics}

The system provides a \textit{Forecasting Engine} and a \textit{Retrospective Engine}. The Forecasting Engine, previously published as DEEP (Discourse Evolution Engine Prediction)~\cite{697729bb748d292568f980b9cc6677a7ddc618a9}, enables journalists to predict future discourse states given historical trajectories and anticipated key events. DEEP models discourse state $\mathcal{S}_t = (\mathbf{V}_t, \mathbf{E}_t, \mathbf{T}_t)$ as a vector capturing volume, emotional intensity across 28 dimensions, and thematic distributions. It uses transformer-based architecture to generate probabilistic forecasts $\mathcal{S}_{t+\Delta} = f(\mathcal{H}_t, \mathcal{K}_{t:t+\Delta}, \boldsymbol{\theta})$ where $\mathcal{H}_t$ represents historical discourse states and $\mathcal{K}_{t:t+\Delta}$ denotes journalist-identified key events within the prediction window. The model outputs Student-t distributions over future states, providing both point estimates and uncertainty quantification to support informed editorial decisions.

{
\textit{This paper focuses on the Retrospective Engine, or REAR (Retrospective Engine for Analysis \& Reasoning), which enables journalists to analyze historical patterns in social movement discourse}. While DEEP addresses forward-looking questions about discourse evolution, REAR supports investigative analysis of past dynamics as requested by our journalism panel. The \#MeToo and BLM cases demonstrate the REAR's capabilities and practical utility for journalists studying SMs.
}

\section{Results on \#MeToo \& Black Lives Matter}
\label{sec:case_studies}

We examined how \#MeToo and Black Lives Matter (BLM) discourse evolved across News and Reddit with respect to 36 Key Political Events (KPEs). We defined KPEs as politically significant developments in the US from September 2024 to August 2025. These events were selected by our journalist co-authors from a well-known website\footnote{\url{www.onthisday.com}}. The 36 KPEs are shown in Appendix (Table~\ref{tab:KEEs}). We investigated three dimensions of SM: discourse volume patterns and their temporal relationship to KPEs, emotion dynamics surrounding these events, and event-specific heterogeneity in SM responses. 

\begin{figure}[t]
  \centering
  \begin{subfigure}{\linewidth}
    \centering
    \includegraphics[width=\textwidth]{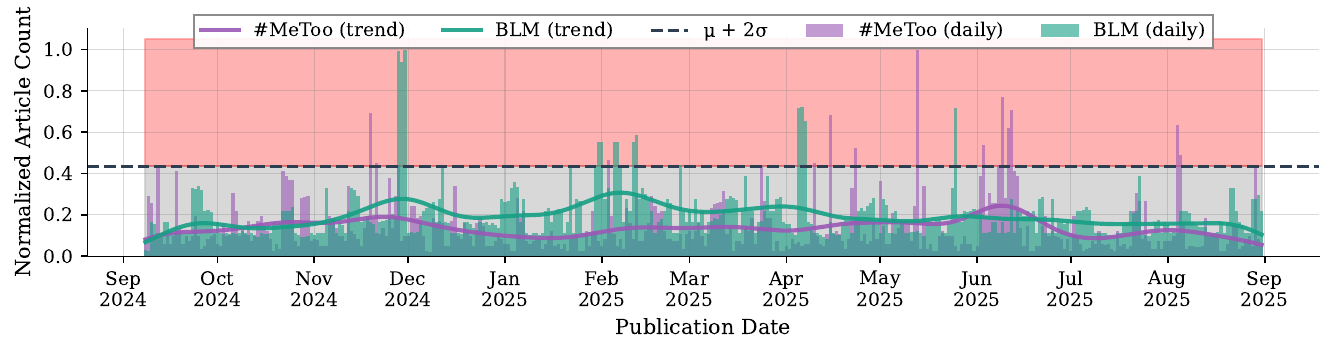}
    \caption{News media}
    \label{fig:participant_accuracy}
  \end{subfigure}
  \hfill
  \begin{subfigure}{\linewidth}
    \centering
    \includegraphics[width=\textwidth]{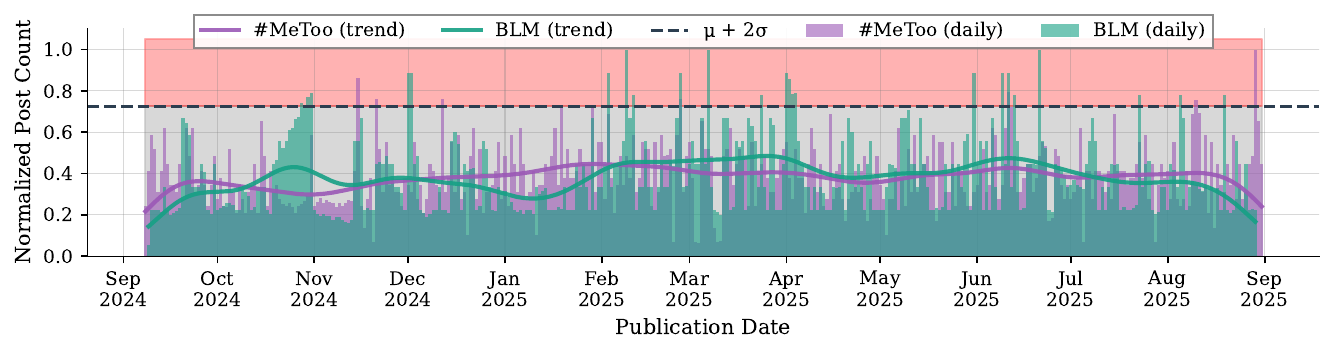}
    \caption{Reddit}
    \label{fig:video_accuracy}
  \end{subfigure}
  \caption{Normalized counts for \#MeToo (purple) and BLM (teal) on (a) news and (b) Reddit. Threshold ($\mu + 2\sigma$, dashed) separates below-threshold (gray) and above-threshold (red) periods. Kernel density curves show overall trends.}
  \label{fig:timedistribution}
\end{figure}

\subsection{Volume \& Temporal Dynamics}

Figure~\ref{fig:timedistribution} displays the temporal distribution of discourse volume across both SMs and platforms, revealing distinct patterns of activity concentration. The visualization shows normalized daily article and post counts alongside a threshold demarcating high-activity periods (defined as exceeding $\mu + 2\sigma$). The pronounced spikes in the time series may be related to KPEs. In the analyses that follow, we systematically test this relationship, examining whether and how KPEs shape SM discourse patterns across news media and Reddit.

\begin{hyp}
    \rule{0.73\columnwidth}{0.5pt} 
    
    \noindent Social movement discourse volume in News media and Reddit exhibits significantly higher daily activity during key political event (KPE) windows compared to non-KPE periods.
    
    \noindent\rule{\columnwidth}{0.5pt}
\end{hyp}

\subsubsection*{Methods}
Discourse volume was measured as the daily count of documents across news media and Reddit. KPE windows were symmetric intervals spanning $k$ days before and after each event date, $k \in \{1, 3, 5, 7, 10\}$. Control periods were matched windows sampled from non-event periods with similar day-of-week distributions.

For each event, we calculated the difference between median daily volume during the event window and its matched control. The aggregate test statistic was the mean difference across all KPEs. Statistical significance was assessed via a permutation test with 10,000 permutations. Two-tailed p-values were computed, and Benjamini-Hochberg FDR correction ($\alpha = 0.05$) for multiple hypothesis testing was applied across window sizes. Effect sizes were quantified using Cohen's d (standardized mean difference), with 95\% bootstrap confidence intervals computed via 1,000 iterations. 

\subsubsection*{Results} 
Figure~\ref{fig:timedistribution} displays effect sizes and 95\% confidence intervals across platforms and window sizes. 

\begin{figure}[t]
    \centering
  \begin{subfigure}[b]{.23\textwidth}
    \centering
    \includegraphics[width=\textwidth]{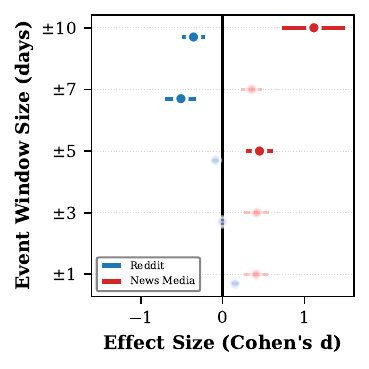}
    \caption{MeToo}
    \label{fig:participant_accuracy}
  \end{subfigure}
  \hfill
  \begin{subfigure}[b]{.23\textwidth}
    \centering
    \includegraphics[width=\textwidth]{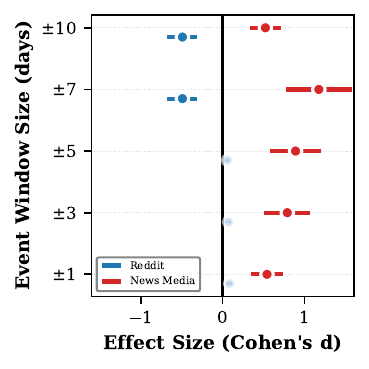}
    \caption{BLM}
    \label{fig:video_accuracy}
  \end{subfigure}
  \caption{Platform-specific effects of key events on discourse volume. Effect sizes (Cohen's d) comparing event windows ($\pm k$ days) versus matched control periods. Darker points denote significant effects ($\alpha = 0.05$); error bars show 95\% CIs.}
  \label{fig:timedistribution}
\end{figure}

\emph{For BLM News coverage, all window sizes showed significant increases during KPE windows.} The $\pm$7-day window exhibited the largest effect (Cohen's $d = 1.17$, $p = 0.001$, 28.53\% increase), with medium-to-large effects persisting across $\pm$3, $\pm$5, and $\pm$10-day windows ($d = 0.79$--0.89, all $p \leq 0.01$). Even the $\pm$1-day window showed a significant small-to-medium effect ($d = 0.54$, $p = 0.018$). All findings remained significant after FDR correction.

\emph{In contrast, BLM discourse on Reddit showed no increases during KPE windows.} Smaller windows ($\pm$1, $\pm$3, $\pm$5 days) produced negligible, non-significant effects ($d = 0.06$--0.08, $p > 0.65$). Longer windows ($\pm$7, $\pm$10 days) revealed significant decreases in discourse volume during event periods ($d = -0.49$, $p = 0.001$ for both, representing 8.5--8.8\% reductions).

\emph{For \#MeToo news coverage, results were more heterogeneous.} The $\pm$10-day window showed a large, significant effect ($d = 1.11$, $p = 0.001$, 23.77\% increase), and the $\pm$5-day window exhibited a medium effect ($d = 0.45$, $p = 0.009$). Shorter windows ($\pm$1, $\pm$3 days) and the $\pm$7-day window showed small-to-medium effects that were not significant after FDR correction ($d = 0.35$--0.41, $p > 0.05$).

\emph{Like BLM, \#MeToo Reddit discourse showed no event-driven increases.} Small windows yielded null effects ($d = -0.006$ to 0.15, $p > 0.49$), while longer windows ($\pm$7, $\pm$10 days) again showed significant decreases ($d = -0.36$ to $-0.51$, $p = 0.001$, representing 6.6--9.5\% reductions).

These results provide mixed support for the hypothesis. While News media consistently shows elevated discourse during KPE windows, Reddit exhibits the opposite pattern, with significant decreases during longer event windows. This platform divergence suggests that key political events trigger different mechanisms: news media amplifies coverage around major events, while Reddit communities may reduce activity during peak news cycles as far as \#MeToo and BLM are concerned. 

\begin{hyp}
    \rule{0.73\columnwidth}{0.5pt} 
    
    \noindent The temporal relationship between key political events (KPEs) and discourse volume exhibits an anticipatory pattern, with pre-event activity significantly exceeding post-event activity.
    
    \noindent\rule{\columnwidth}{0.5pt}
\end{hyp}

\subsubsection*{Methods}
For each event, we extracted symmetric windows spanning $\pm k$ days, $k =7$ days. The event day itself was excluded to isolate anticipatory versus reactive patterns. Pre-event volume was calculated as the mean daily count across days $t \in [-7, -1]$ relative to the event, while post-event volume was the mean across days $t \in [+1, +7]$. To control for baseline trends, we normalized both periods by a reference window at days $t \in [-14, -8]$. The test statistic was the mean difference (pre minus post) across all events, where positive (resp. negative) values indicate anticipatory (resp. reactive) patterns.
Statistical significance was assessed via a permutation test with 10,000 iterations, randomly reassigning event dates to alternative time points while preserving window structure and temporal constraints. Two-tailed p-values were computed comparing the observed mean difference to the null distribution. We also calculated directional (one-tailed) p-values testing the specific hypothesis that pre-event volume exceeds post-event volume. Benjamini-Hochberg FDR correction ($\alpha = 0.05$) for multiple hypothesis testing was applied. Effect sizes were quantified using Cohen's d, and 95\% confidence intervals were constructed via bootstrap resampling with 1,000 iterations.

\subsubsection*{Results}

Contrary to the hypothesis, we found no significant temporal precedence of pre-event over post-event discourse volume across any platform or SM. Table~\ref{tab:event_study} presents the full results for the ±7 day window analysis.

\begin{table}[t]
\footnotesize
\centering
\caption{Pre-Event vs. Post-Event Discourse Volume}
\label{tab:event_study}
\renewcommand{\arraystretch}{0.9}
\begin{tabularx}{\columnwidth}{l>{\RaggedRight}X>{\RaggedRight}X>{\RaggedRight}X>{\RaggedRight}X}
\toprule
\multirow{2}{*}{\textbf{Metric}} & \multicolumn{2}{c|}{\textbf{News media}} & \multicolumn{2}{c}{\textbf{Reddit}} \\
\cline{2-5}
 & \textit{BLM} & \textit{\#MeToo} & \textit{BLM} & \textit{\#MeToo} \\
\midrule
Mean Diff. & $+0.48$ & $-0.52$ & $-11.34$ & $-19.35$ \\
95\% CI & $[-2.75, 3.62]$ & $[-4.41, 3.55]$ & $[-34.78, 7.80]$ & $[-65.99, 18.08]$ \\
Cohen's $d$ & $0.055$ & $-0.041$ & $-0.162$ & $-0.140$ \\
$p$-value & 0.770 & 0.804 & 0.240 & 0.330 \\
\bottomrule
\end{tabularx}
\begin{tablenotes}
\footnotesize
\item Note: Mean Diff. indicates differences between the pre- and post-event windows; $p$-values are FDR-corrected.
\end{tablenotes}
\end{table}

None of the four datasets exhibited statistically significant temporal patterns. Mean differences ranged from $+0.48$ (BLM news, slight anticipatory trend) to $-19.35$ (\#MeToo Reddit, slight reactive trend), but all remained non-significant with p-values ranging from 0.240 to 0.804. The proportion of events showing either anticipatory or reactive patterns was nearly balanced across datasets, ranging from 47.2\% to 55.6\%. Effect sizes were uniformly small (Cohen's $|d|$ $<$ 0.17), and all 95\% confidence intervals included zero, indicating substantial uncertainty in the direction and magnitude of any temporal effects.

Figure~\ref{fig:pre_post_scatter} in the Appendix shows event-level heterogeneity underlying these aggregate null results. Individual events varied considerably in their temporal patterns, with roughly half showing anticipatory patterns (points above the diagonal) and half showing reactive patterns (points below the diagonal). \emph{This heterogeneity suggests that temporal dynamics may be event-specific rather than reflecting a consistent movement-level pattern.}

\begin{takeaway}
    \textbf{Takeaways (Volume \& Temporal Dynamics) } 
    \begin{enumerate}[leftmargin=*, labelindent=0pt, itemsep=3pt]
        \item News coverage volume is significantly higher during key political event (KPE) windows, while Reddit activity is significantly lower during the same periods.
        \item BLM shows stronger and more consistent news media responses to KPEs compared to \#MeToo.
        \item No universal temporal pattern emerges: individual events show mixed anticipatory vs. reactive dynamics with no consistent directionality.
    \end{enumerate}
\end{takeaway}

\subsection{Event-Specific Analysis}
While aggregate analyses revealed platform-level differences in how News media and Reddit responded to Key Political Events, these patterns may obscure variations across individual events. Not all KPEs necessarily exert equal influence on SM discourse. We examined this heterogeneity via an event-specific analysis.

\begin{hyp}
    \rule{0.73\columnwidth}{0.5pt} 
    
    \noindent Individual key political events (KPEs) vary significantly in their impact on SM discourse volume, with some events triggering substantial increases, others causing decreases, and many showing no significant effect relative to baseline periods.
    
    \noindent\rule{\columnwidth}{0.5pt}
\end{hyp}

\subsubsection*{Methods}
For each event, we defined event windows as symmetric intervals spanning $\pm k$ days around the event date, where $k=7$. Inside-window volume was calculated as the mean daily count within the event's $\pm k$ window. Baseline volume was computed as the mean daily count across all days not falling within any event window (for all events at the given window size $k$), providing a global reference level excluding all event-related periods. Statistical significance was assessed via a permutation test with 1,000 iterations per event-window combination. For each permutation, we randomly sampled $n$ days from the complete time series (where $n$ equals the event window size) and compared their mean volume to the mean of all remaining days, generating a null distribution under the hypothesis that the observed event window represents a random sample of days rather than a period with event-specific effects. Two-tailed p-values were computed as the proportion of permutations where the absolute difference equaled or exceeded the observed absolute difference, allowing detection of both volume increases and decreases. Benjamini-Hochberg FDR correction ($\alpha = 0.05$) for multiple hypothesis testing was applied across all event-window tests to control for multiple comparisons. 

\subsubsection*{Results}
Table~\ref{tab:kpe} presents percentage differences in discourse volume between event windows ($\pm 7$ days) and baseline periods for each KPE across news media and Reddit. \emph{The results reveal substantial heterogeneity in event impacts, with effects varying in both magnitude and direction. \#MeToo and BLM exhibited markedly different patterns in news coverage but strikingly similar patterns on Reddit.}

\begin{table}[t]
\centering
\caption{Key Political Events Analysis}
\label{tab:kpe}
\renewcommand{\arraystretch}{0.73}
\begin{tabularx}{\columnwidth}{r|>{\Centering}X>{\Centering}X|>{\Centering}X>{\Centering}X}
\toprule
\multirow{2}{*}{\textbf{KPE}} & \multicolumn{2}{c|}{\textbf{MeToo}} & \multicolumn{2}{c}{\textbf{BLM}} \\
\cline{2-5}
 & \textit{News} & \textit{Reddit} & \textit{News} & \textit{Reddit} \\
\midrule
Sep 15, 2024    & \cellcolor{red!2} -2.0    & \cellcolor{red!0} -0.1    & \cellcolor{red!9} -8.9    & \cellcolor{green!1} 1.0 \\
Sep 24, 2024    & \cellcolor{green!24} 24.4 & \cellcolor{green!1} 0.5   & \cellcolor{red!21} -20.9  & \cellcolor{red!0} -0.2 \\
Oct 1, 2024     & \cellcolor{green!23} 22.9 & \cellcolor{red!7} -6.5    & \cellcolor{red!24} -24.4  & \cellcolor{red!6} -5.7 \\
Oct 2, 2024     & \cellcolor{green!15} 15.0 & \cellcolor{red!7} -7.2    & \cellcolor{red!24} -23.7  & \cellcolor{red!5} -4.9 \\
Nov 5, 2024     & \cellcolor{green!5} 4.5   & \cellcolor{red!86} -85.8*** & \cellcolor{green!55} 54.7*** & \cellcolor{red!87} -86.8*** \\
Nov 12, 2024    & \cellcolor{green!46} 46.0*** & \cellcolor{red!58} -58.4*** & \cellcolor{green!58} 58.2*** & \cellcolor{red!59} -58.7*** \\
Nov 14, 2024    & \cellcolor{green!41} 41.2*  & \cellcolor{red!47} -47.4*** & \cellcolor{green!40} 39.9**  & \cellcolor{red!49} -49.1*** \\
Nov 25, 2024    & \cellcolor{green!46} 46.0*** & \cellcolor{red!2} -2.1   & \cellcolor{red!1} -0.5    & \cellcolor{red!3} -3.3 \\
Jan 4, 2025     & \cellcolor{red!11} -10.5  & \cellcolor{green!3} 2.9   & \cellcolor{green!13} 13.0 & \cellcolor{green!11} 10.6 \\
Jan 10, 2025    & \cellcolor{red!14} -14.2  & \cellcolor{green!5} 5.2   & \cellcolor{green!7} 6.9   & \cellcolor{green!13} 13.1 \\
Jan 20, 2025    & \cellcolor{green!7} 7.1   & \cellcolor{green!9} 9.0   & \cellcolor{green!43} 43.2*** & \cellcolor{green!19} 18.6 \\
Jan 24, 2025    & \cellcolor{red!3} -2.6    & \cellcolor{green!9} 8.8   & \cellcolor{green!56} 55.7*** & \cellcolor{green!20} 20.4 \\
Jan 27, 2025    & \cellcolor{red!3} -2.5    & \cellcolor{green!9} 8.9   & \cellcolor{green!60} 59.5*** & \cellcolor{green!19} 18.7 \\
Jan 30, 2025    & \cellcolor{red!4} -4.4    & \cellcolor{green!6} 6.4   & \cellcolor{green!52} 51.6*** & \cellcolor{green!13} 13.3 \\
Jan 31, 2025    & \cellcolor{red!13} -13.3  & \cellcolor{green!7} 7.1   & \cellcolor{green!53} 52.9*** & \cellcolor{green!13} 13.0 \\
Feb 5, 2025     & \cellcolor{red!16} -15.7  & \cellcolor{green!7} 6.6   & \cellcolor{green!53} 53.4*** & \cellcolor{green!7} 7.1 \\
Feb 6, 2025     & \cellcolor{red!12} -11.8  & \cellcolor{green!8} 7.5   & \cellcolor{green!55} 54.9*** & \cellcolor{green!6} 6.1 \\
Feb 12, 2025    & \cellcolor{red!1} -0.8    & \cellcolor{green!5} 4.9   & \cellcolor{green!49} 48.6*** & \cellcolor{green!6} 6.1 \\
Feb 13, 2025    & \cellcolor{green!2} 2.4   & \cellcolor{green!5} 5.1   & \cellcolor{green!48} 48.1*** & \cellcolor{green!8} 7.7 \\
Feb 18, 2025    & \cellcolor{green!4} 4.3   & \cellcolor{red!6} -5.6    & \cellcolor{green!43} 42.5**  & \cellcolor{red!2} -2.3 \\
Feb 20, 2025    & \cellcolor{green!2} 2.4   & \cellcolor{red!6} -5.5    & \cellcolor{green!44} 43.7**  & \cellcolor{red!2} -1.9 \\
Feb 21, 2025    & \cellcolor{red!0} -0.4    & \cellcolor{red!6} -6.0    & \cellcolor{green!40} 40.2**  & \cellcolor{red!1} -1.1 \\
Feb 28, 2025    & \cellcolor{red!11} -11.0  & \cellcolor{red!10} -10.0  & \cellcolor{green!43} 43.0**  & \cellcolor{red!6} -5.8 \\
Mar 3, 2025     & \cellcolor{red!22} -22.3  & \cellcolor{red!22} -21.6  & \cellcolor{green!22} 21.6 & \cellcolor{red!16} -15.6 \\
Apr 2, 2025     & \cellcolor{red!17} -16.7  & \cellcolor{green!12} 12.2 & \cellcolor{green!26} 25.7 & \cellcolor{green!11} 11.2 \\
Apr 9, 2025     & \cellcolor{green!21} 21.0 & \cellcolor{green!17} 16.6 & \cellcolor{green!41} 40.7**  & \cellcolor{green!12} 12.3 \\
May 4, 2025     & \cellcolor{red!3} -3.3    & \cellcolor{green!13} 12.7 & \cellcolor{green!18} 18.1 & \cellcolor{green!17} 16.8 \\
Jun 6, 2025     & \cellcolor{green!78} 78.1*** & \cellcolor{green!25} 24.9 & \cellcolor{green!28} 28.0 & \cellcolor{green!14} 14.0 \\
Jun 7, 2025     & \cellcolor{green!72} 71.7*** & \cellcolor{green!25} 24.8 & \cellcolor{green!30} 29.5 & \cellcolor{green!15} 14.5 \\
Jun 9, 2025     & \cellcolor{green!71} 71.2*** & \cellcolor{green!25} 25.0 & \cellcolor{green!29} 28.7 & \cellcolor{green!16} 16.4 \\
Jun 14, 2025    & \cellcolor{green!57} 57.1*** & \cellcolor{green!21} 21.2 & \cellcolor{green!24} 24.4 & \cellcolor{green!17} 17.1 \\
Jul 4, 2025     & \cellcolor{green!37} 37.3 & \cellcolor{green!15} 14.9 & \cellcolor{red!13} -12.7  & \cellcolor{green!3} 2.7 \\
Aug 12, 2025    & \cellcolor{green!5} 5.0   & \cellcolor{green!11} 10.5 & \cellcolor{red!14} -13.5  & \cellcolor{green!0} 0.2 \\
Aug 15, 2025    & \cellcolor{green!10} 10.0 & \cellcolor{green!8} 8.0   & \cellcolor{red!18} -18.3  & \cellcolor{red!2} -2.1 \\
Aug 18, 2025    & \cellcolor{green!18} 17.6 & \cellcolor{green!16} 15.9 & \cellcolor{red!15} -14.5  & \cellcolor{green!1} 1.2 \\
Aug 22, 2025    & \cellcolor{green!27} 27.3 & \cellcolor{green!21} 21.4 & \cellcolor{green!20} 19.6 & \cellcolor{green!11} 10.8 \\
\bottomrule
\end{tabularx}
\begin{tablenotes}
\footnotesize
\item Note: *** $p < 0.001$, ** $p < 0.01$, * $p < 0.05$. Cell colors indicate magnitude: green for positive values, red for negative values, with intensity proportional to absolute value.
\end{tablenotes}
\end{table}

\emph{\#MeToo and BLM discourse diverged substantially on News.} \#MeToo events showed dramatic increases in mid-2025, coinciding with immigration actions: June 6, 2025 (ICE sweeps in Los Angeles) exhibited a 78.1\% increase (p < 0.001), while June 7 (deployment of 2,000 National Guard troops) and June 9 (deployment of 700 Marines) showed 71.7\% and 71.2\% increases respectively (both p < 0.001). The 'No Kings' protests across all 50 states on June 14 saw elevated coverage at 57.1\% (p < 0.001). Events in November 2024 also produced sustained elevated coverage: the announcement of Elon Musk and Vivek Ramaswamy heading the Department of Government Efficiency (Nov 12, 46.0\%, p < 0.001), Republican House control (Nov 14, 41.2\%, p < 0.05), and dismissal of charges against President Trump (Nov 25, 46.0\%, p < 0.001). 

In contrast, BLM news coverage showed elevated activity primarily from President Trump's (Jan 20 2025) inauguration through February 2025. The inauguration itself (Jan 20) triggered a 43.2\% increase (p < 0.001), followed by sustained elevated coverage through the firing of prosecutors (Jan 24, 55.7\%, p < 0.001; Jan 27, 59.5\%, p < 0.001), the sentencing of Senator Menendez (Jan 30, 51.6\%, p < 0.001), the firing of January 6 event investigators (Jan 31, 52.9\%, p < 0.001), and executive actions including the transgender sports ban (Feb 5, 53.4\%, p < 0.001) and Gaza development comments (Feb 6, 54.9\%, p < 0.001). This pattern extended through cabinet confirmations (Tulsi Gabbard on Feb 12, 48.6\%, p < 0.001; Robert F. Kennedy Jr. on Feb 13, 48.1\%, p < 0.001) and foreign policy developments (US-Russia talks on Feb 18, 42.5\%, p < 0.01; Kash Patel's FBI confirmation on Feb 20, 43.7\%, p < 0.01; firing of Joint Chiefs Chairman on Feb 21, 40.2\%, p < 0.01; Trump-Zelensky argument on Feb 28, 43.0\%, p < 0.01). Later events like tariff reversals (Apr 9, 40.7\%, p < 0.01) also showed significant increases.

\emph{
On Reddit, however, both SMs exhibited remarkably similar patterns, suggesting that major political events may have crowded out SM discourse on Reddit.} For both \#MeToo and BLM, President Trump's election victory on November 5 produced the largest reductions (-85.8\% and -86.8\% respectively, both p < 0.001), followed by the DOGE announcement on November 12 (-58.4\% and -58.7\%, both p < 0.001) and Republican House control on November 14 (-47.4\% and -49.1\%, both p < 0.001). This pattern suggests that the post-election political transition dominated Reddit discourse, displacing discussion of both SMs.

Overall, 30 out of a total of 144 total event-platform combinations (20.8\%) showed statistically significant effects after FDR correction. These results support the hypothesis that individual KPEs vary significantly in their discourse impacts, with substantial heterogeneity across events, platforms, and movements.

\begin{takeaway}
    \textbf{Takeaways (Event-Specific Analysis) } 
    \begin{enumerate}[leftmargin=*, labelindent=0pt, itemsep=3pt]
        \item Individual KPEs vary in their impact, with 21\% of event-platform combinations showing significant effects on discourse volume.
        \item \#MeToo news coverage peaked during 2025 immigration enforcement actions, while BLM surged during President Trump's inauguration and early administration actions.
        \item Reddit activity consistently decreased across both SMs during the November 2024 election period, suggesting platform-wide attention displacement.
    \end{enumerate}
\end{takeaway}

\subsection{Emotional Dynamics}
Beyond discourse volume, emotions constitute a central dimension of social movement discourse. We examine whether key political events systematically affect emotional intensity in movement discourse, and whether emotions exhibit anticipatory or reactive temporal patterns around KPEs.

\begin{hyp}
    \rule{0.73\columnwidth}{0.5pt} 
    
    \noindent Emotion intensity differs significantly in temporal windows surrounding key events compared to baseline periods.
    
    \noindent\rule{\columnwidth}{0.5pt}
\end{hyp}

\subsubsection*{Methods}

Emotion intensity for each content item (post or article) was measured as the sum of emotion probabilities across all non-neutral emotion categories from the GoEmotions dataset \cite{DBLP:conf/acl/DemszkyMKCNR20}.  This measure was aggregated daily by computing the mean intensity across all content published that day.

We define event windows as symmetric intervals spanning $k$ days before through $k$ days after each KPE date, where $k \in \{1, 3, 5, 7, 10\}$ days. Days within event windows were compared against baseline periods, defined as all days falling outside any event window for the given window size $k$. To prevent contamination, we excluded days within a $\pm 14$-day buffer around events from the baseline.

For each window size, we compared baseline versus event-window emotion intensity distributions using Mann-Whitney U tests. Effect sizes were quantified using Cohen's $d$ (standardized mean difference), with 95\% bootstrap confidence intervals computed via 2,000 iterations. Statistical significance was assessed via Benjamini-Hochberg FDR correction ($\alpha = 0.05$).

\subsubsection*{Results}

\begin{figure}[t]
    \centering
  \begin{subfigure}[b]{.21\textwidth}
    \centering
    \includegraphics[width=\textwidth]{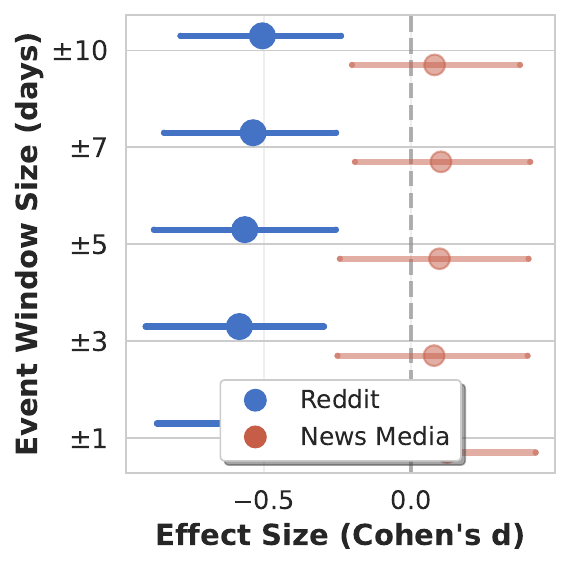}
    \caption{MeToo \label{fig:emotion_fp_a}}
    
  \end{subfigure}
  \hfill
  \begin{subfigure}[b]{.21\textwidth}
    \centering
    \includegraphics[width=\textwidth]{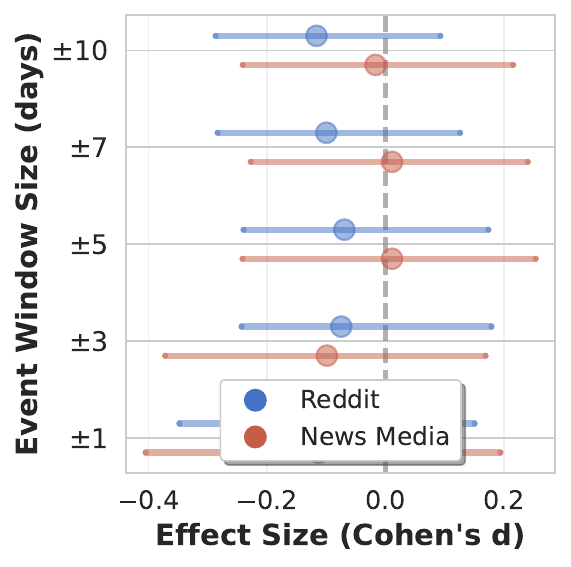}
    \caption{BLM \label{fig:emotion_fp_b}}
    
  \end{subfigure}

  \begin{subfigure}[b]{.23\textwidth}
    \centering
    \includegraphics[width=\textwidth]{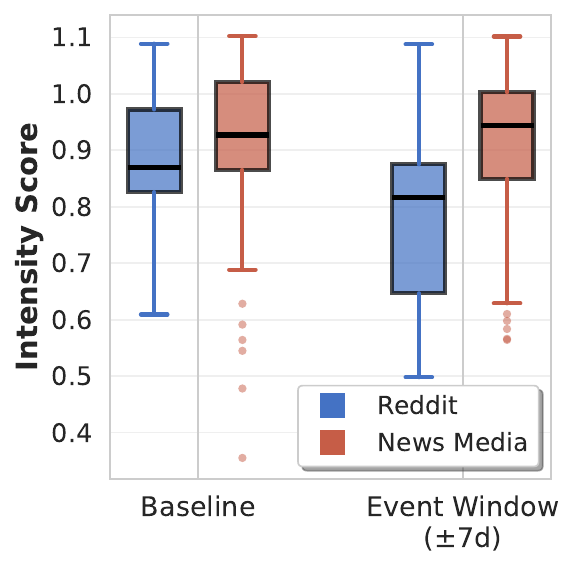}
    \caption{MeToo \label{fig:emotion_fp_c}}
    
  \end{subfigure}
  \hfill
  \begin{subfigure}[b]{.23\textwidth}
    \centering
    \includegraphics[width=\textwidth]{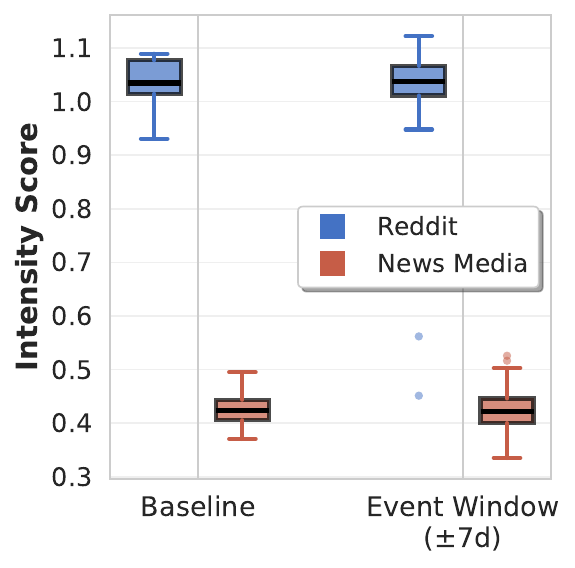}
    \caption{BLM \label{fig:emotion_fp_d}}
    
  \end{subfigure}
  
    \caption{Key event effects on emotion intensity. (a-b) Cohen's d comparing event windows ($\pm k$ days) versus baseline. Darker points: FDR-significant ($\alpha = 0.05$); error bars: 95\% CIs. (c-d) Emotion intensity distributions for baseline versus $\pm$7-day windows.}
  \label{fig:emotion_fp}
\end{figure}

Figure~\ref{fig:emotion_fp} displays effect sizes and intensity distributions across SMs, platforms and window sizes. Unlike the hypothesis that emotion intensity increases around key political events, we found no significant increases in any SM-platform combination. Instead, several analyses revealed significant decreases in emotional expression during event windows.

\emph{For \#MeToo on Reddit, all window sizes consistently showed reduced emotion intensity during event periods.} For example, the $\pm$7-day window exhibited Cohen's $d = -0.51$, FDR-corrected $p = 2.53 \times 10^{-4}$. All findings remained significant after FDR correction. As shown in Figure~\ref{fig:emotion_fp_c}, the comparison of distributions reveals that \#MeToo discourse on Reddit becomes notably less emotional during event windows compared to baseline periods.

In contrast, \#MeToo News coverage showed no significant changes in emotion intensity across any window size ($d \in [-0.25, 0.65]$, all $p > 0.05$). \emph{Interestingly, Figure~\ref{fig:emotion_fp_c} shows that news articles exhibit substantially higher baseline emotion intensity than Reddit posts, suggesting that journalistic coverage of \#MeToo seems more emotional, regardless of political events.}

For BLM, neither platform showed significant emotion intensity changes during event windows. For example, BLM news coverage produced negligible effects across all window sizes ($d \in [-0.4, 0.25]$, all $p > 0.65$). Figure~\ref{fig:emotion_fp_d} demonstrates that BLM Reddit posts exhibit higher baseline emotion intensity than news articles—the opposite pattern observed for \#MeToo—suggesting SMs differ in how emotions are expressed across platforms.

\emph{
These results provide no support for the hypothesis that emotion intensity increases around key political events.}\newpage  

\begin{hyp}
    \rule{0.73\columnwidth}{0.5pt} 
    
    \noindent Emotion intensity exhibits anticipatory patterns around key political events, with pre-event intensity significantly exceeding post-event intensity.
    
    \noindent\rule{\columnwidth}{0.5pt}
\end{hyp}

\subsubsection*{Methods}
To examine temporal patterns of emotional expression around KPEs, we conducted event-level pre and post comparisons of emotion intensity. Each of the 36 KPEs was manually categorized into one of three types based on their primary policy domain: (1) \textit{Domestic Policy} (e.g., executive orders, cabinet confirmations), (2) \textit{Elections} (e.g., election results, transitions of power), and (3) \textit{Foreign Policy} (e.g., international negotiations, diplomatic meetings). The full event-category mapping is in Appendix (Table~\ref{tab:KEEs}). Emotion intensity was measured as described in the previous section, representing the sum of all non-neutral emotion probabilities for each content item, aggregated daily. For each KPE, we defined symmetric temporal windows: a pre-event period spanning days $t \in [-7, -1]$ relative to the event date, and a post-event period spanning days $t \in [+1, +7]$. The event day itself ($t = 0$) was excluded to isolate anticipatory versus reactive patterns.

For each individual event, we extracted all content items (posts or articles) published during its pre-event and post-event windows. We then compared the distribution of daily mean intensity scores between these two periods using a two-sided Mann-Whitney U test. This yielded one p-value per event, testing whether pre-event and post-event intensity distributions differed significantly for that specific event. Benjamini-Hochberg FDR correction ($\alpha = 0.05$) was applied across all 36 events within each movement$\times$platform combination to control for multiple comparisons.

\subsubsection*{Results}

\begin{figure}[t]
    \centering

  \begin{subfigure}[b]{.23\textwidth}
    \centering
    \includegraphics[width=\textwidth]{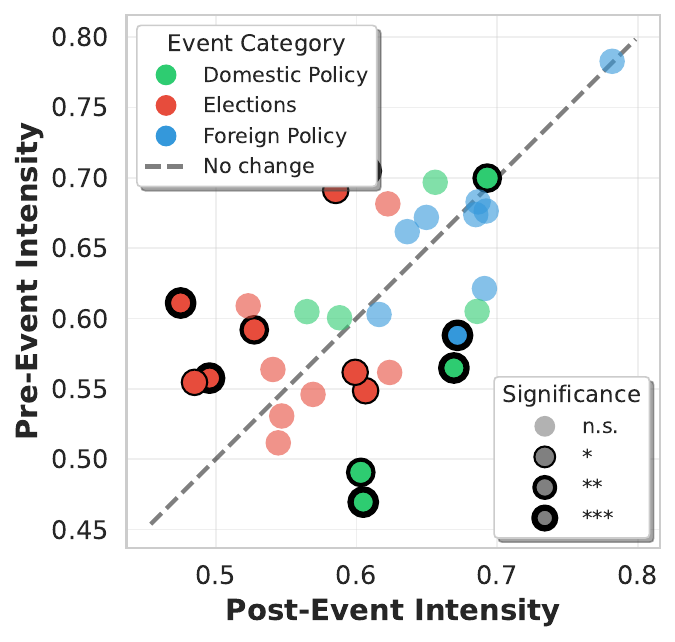}
    \caption{MeToo (News) \label{fig:emotion_scatter_a}}
    
  \end{subfigure}
  \hfill
  \begin{subfigure}[b]{.23\textwidth}
    \centering
    \includegraphics[width=\textwidth]{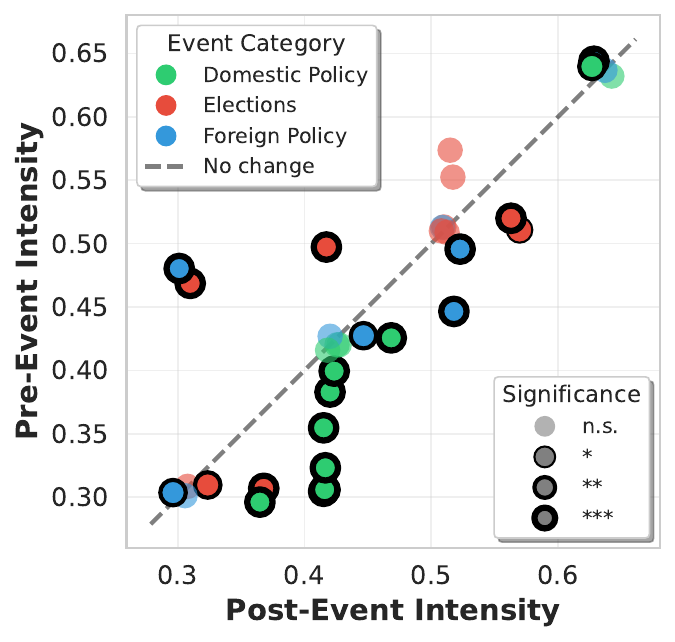}
    \caption{MeToo (Reddit) \label{fig:emotion_scatter_b}}
    
  \end{subfigure}

  \begin{subfigure}[b]{.23\textwidth}
    \centering
    \includegraphics[width=\textwidth]{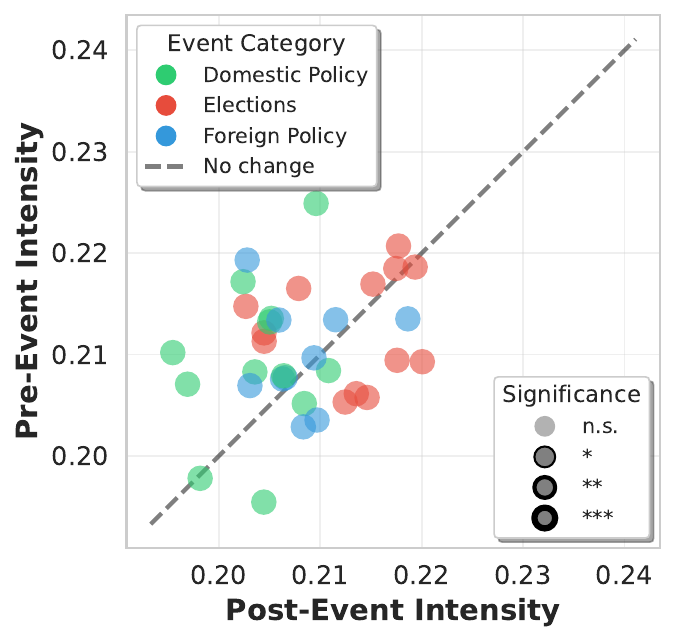}
    \caption{BLM (News) \label{fig:emotion_scatter_c}}
    
  \end{subfigure}
  \hfill
  \begin{subfigure}[b]{.23\textwidth}
    \centering
    \includegraphics[width=\textwidth]{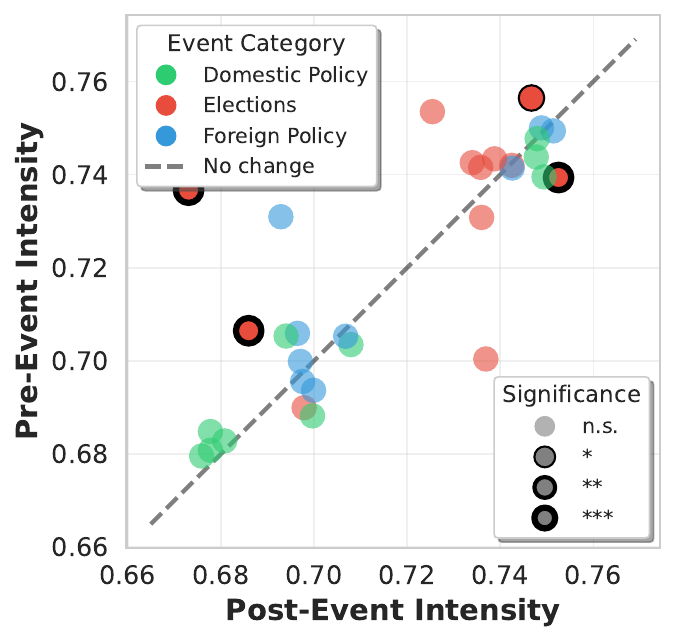}
    \caption{BLM (Reddit) \label{fig:emotion_scatter_d}}
    
  \end{subfigure}
  
\caption{Pre- versus post-event emotion intensity for KPEs by category: Domestic Policy (green), Elections (red), Foreign Policy (blue). Points above diagonal: anticipatory patterns; below: reactive patterns. Borders: FDR-significant effects.}
\label{fig:emotion_scatter}
 
\end{figure}

Figure~\ref{fig:emotion_scatter} shows pre-event versus post-event emotion intensity for all 36 KPEs across movement-platform combinations. Contrary to the hypothesis that emotion intensity follows consistent anticipatory patterns, our results reveal substantial event-level heterogeneity with no universal temporal pattern.

\emph{For \#MeToo news coverage (Figure~\ref{fig:emotion_scatter_a}), several individual events showed significant pre-post differences, but these varied in direction.} Events related to domestic policy and elections exhibited both anticipatory and reactive patterns, with no consistent trend within event categories. Notably, no foreign policy events showed significant intensity changes, suggesting that international developments may have limited relevance to \#MeToo discourse in news media.

In contrast, \#MeToo on Reddit (Figure~\ref{fig:emotion_scatter_b}) exhibited the most extensive temporal dynamics, with 20 out of 36 events (57.1\%) showing significant pre-post differences after FDR correction. This pattern aligns with our previous finding (Hypothesis 3) that \#MeToo Reddit discourse exhibits elevated emotional intensity around KPEs overall. Remarkably, domestic policy events showed a consistent reactive pattern, with all significant domestic policy events falling below the diagonal, indicating heightened emotional expression \textit{after} rather than before these events. \emph{This suggests that domestic policy developments trigger emotional responses in Reddit \#MeToo discourse rather than building anticipatory tension.}

\emph{For BLM news coverage (Figure~\ref{fig:emotion_scatter_c}), no events exhibited statistically significant pre-post intensity differences.} This null finding indicates that BLM-related news coverage maintains relatively stable emotional intensity regardless of KPEs timeline.

BLM on Reddit (Figure~\ref{fig:emotion_scatter_d}) showed significant pre-post differences for only 4 events (11\%), all of which belonged to the elections category. Three of these four election-related events exhibited anticipatory patterns (above the diagonal), indicating that election outcomes and transitions of power triggered heightened emotional expression in BLM Reddit discourse before they occurred. 

These results provide no support for the hypothesis that emotion intensity consistently exhibits anticipatory patterns. Instead, temporal dynamics appear highly event-specific and platform-dependent. 

\begin{takeaway}
    \textbf{Takeaways (Emotional Dynamics) } 
    \begin{enumerate}[leftmargin=*, labelindent=0pt, itemsep=3pt]
        \item Emotion intensity does not increase around KPEs. \#MeToo on Reddit even shows significant decreases. 
        \item News exhibit higher emotion intensity than Reddit for \#MeToo, while the opposite pattern holds for BLM.
        \item Domestic policy events trigger reactive emotional responses in \#MeToo Reddit. Election events produce mixed patterns across movements.

    \end{enumerate}
\end{takeaway}

\section{Discussion and Conclusion}

Our analysis, enabled by SMART, reveals fundamental platform differences in how social movements respond to key political events. News discourse volume increased significantly during KPE windows (Cohen's $d$ up to 1.17 for BLM), while Reddit showed the opposite pattern with decreases during longer event windows (Cohen's $d\in[-0.49,-0.51]$). This divergence suggests that news organizations amplify coverage during major political events, while grassroots discourse may be reduced, possibly due to attention shifting to breaking news. \emph{We caution journalists that social media activity during major political events may not reflect grassroots engagement levels. We recommend monitoring baseline discourse patterns between events as this may better capture movement vitality than tracking activity during peak news cycles.}

Our temporal analysis revealed no consistent anticipatory or reactive patterns in discourse volume, contrary to expectations that movements might mobilize before major events. However, substantial heterogeneity emerged in how individual KPEs affected discourse, with 21\% of event-platform combinations showing significant effects. \#MeToo news coverage surged around 2025 immigration enforcement actions (June 6: +78.1\%, FDR-corrected $p<0.001$), while BLM coverage peaked during President Trump's inauguration and early administration actions. \emph{We recommend that journalists consider SM-specific sensitivities when planning coverage, as not all political events generate equivalent discourse responses across different movements.}

Unexpectedly, emotion intensity did not increase around KPEs. \#MeToo Reddit showed significant emotion decreases during event windows (Cohen $d=-0.51$, FDR-corrected $p<0.001$), while news media maintained stable baseline intensity. This challenges assumptions that politically charged events consistently amplify emotional expression. Event-level analysis revealed context-dependent patterns: \#MeToo Reddit showed reactive emotional responses to domestic policy events, while BLM Reddit showed anticipatory patterns for election-related events. \emph{We recommend that journalists monitor multiple platforms, as emotional trajectories vary based on the type of political development and the movement's core concerns.}

{
Overall, our analysis demonstrates the utility of SMART for tracking and analyzing social movement discourse across multiple data sources. Future work includes analyzing additional movements and platforms and linking discourse patterns to offline outcomes.
}

{
\subsubsection*{Limitations and Ethical Considerations}

Like any study, ours has limitations. Our findings are dependent on the selected 36 political events, two data platforms (Reddit and news media), and two social movements (\#MeToo and BLM)—different selections may yield different patterns and our results may not generalize to other contexts. Additionally, temporal associations between KPEs and discourse dynamics do not establish causality—shifts may reflect media agenda-setting, organic responses, or algorithmic changes.

We followed strict ethical standards, using only publicly available data and no personal information. Recognizing that movement-tracking technologies could be repurposed for surveillance or discrediting activities, we implemented safeguards: (1) restricting SMART access to verified journalists, (2) maintaining human oversight, and (3) releasing only aggregated data.} 

\begin{acks}
This research was supported by the Roberta Buffett Institute for Global Affairs at Northwestern University. Additional support was provided by Amazon Web Services (AWS) through the AWS Cloud Credits for Research program. 
\end{acks}

\newpage


\bibliographystyle{ACM-Reference-Format}
\bibliography{bibfile}


\appendix

\section{SMART Participating Organizations}

To ensure that SMART serves as an effective tool for journalists seeking to understand the evolution and community impact of social movements, we engaged in a comprehensive stakeholder consultation process involving multiple phases of input and feedback.

\noindent \textbf{Initial Stakeholder Interviews.}
We conducted in-depth interviews with eight journalists representing four diverse news organizations: Grist (2 journalists), The 19th (1 journalist), The Associated Press (3 journalists), and The Examination (2 journalists). These interviews served multiple purposes: first, to establish working definitions of “social movement" from the journalistic perspective; second, to identify current methodologies journalists employ for tracking social movements; and third, to document existing pain points and workflow challenges in their reporting processes. Based on these insights, we collaboratively brainstormed potential features and functionalities for the SMART platform.

\noindent \textbf{Prototype Evaluation and Refinement.}
Following the initial development phase, we returned to the same cohort of journalists to demonstrate the SMART prototype. This iterative feedback process allowed us to collect targeted input on the tool's functionality, user interface, and overall efficacy, leading to substantial refinements in the platform's design and capabilities.

\noindent \textbf{Extended Workshop Validation}
To broaden our stakeholder engagement and validate findings across a more diverse journalism ecosystem, we organized a workshop in March 2025\footnote{Reference omitted for the sake of anonymity.}. This workshop included representatives from nine additional organizations, encompassing both traditional and digital media outlets: The Wall Street Journal, American Press Institute, LocalMedia Association, The Washington Post, Borderless Magazine, Institute for Nonprofit News, ChicagoPublicMedia, and Chicago Sun-Times Media.

\begin{table}[t]
\centering
\caption{Summary of Multi-layer Data Extraction Results}
\label{tab:data}
\renewcommand{\arraystretch}{0.7}
\begin{tabularx}{\columnwidth}{r|>{\Centering}X>{\Centering}X|>{\Centering}X>{\Centering}X}
\toprule
\multirow{2}{*}{\textbf{Layer}} & \multicolumn{2}{c|}{\textbf{News media}} & \multicolumn{2}{c}{\textbf{Reddit}} \\
\cline{2-5}
 & \textit{\#MeToo} & \textit{BLM} & \textit{\#MeToo} & \textit{BLM} \\
\midrule
$L_0$ & 3,788 & 819 & 3,826 & 890\\
$L_1$ & 5,933 & 159 & 41,671 & 14,084\\
$L_2$ & 913 & 397 & 53,272 & 12,991\\
$L_3$ & 800 & 1,044 & --- & 24,806\\
$L_4$ & 1,256 & 2,273 & 94,153 & --- \\
$L_5$ & 3,419 & 5,888 & --- & 47,320\\
$L_6$ & 11,914 & 23,381 & 142,719 & 88,392\\
$L_7$ & 37,450 & 46,158 & --- & 172,663\\
$L_8$ & 79,281 & 90,642 & 214,929  & 432,952\\ \midrule 
Ext. & 861,709 & 835,702 & 2,217,355 & 1,973,827 \\ \midrule 
\textbf{Tot. }& \multicolumn{2}{c}{\textbf{1,006,463}} & \multicolumn{2}{c}{\textbf{2,767,925}}\\
\bottomrule
\end{tabularx}

\begin{tablenotes}
    \footnotesize 
    \item Note: ``---'' indicates cases where no documents met the filtering criteria for that specific movement-platform combination. The composition of layers vary by dataset based on keyword co-occurrence patterns.
\end{tablenotes}
\end{table}

    



\begin{figure}[t]
     \centering

     \subfloat[][]{\includegraphics[width=.23\textwidth]{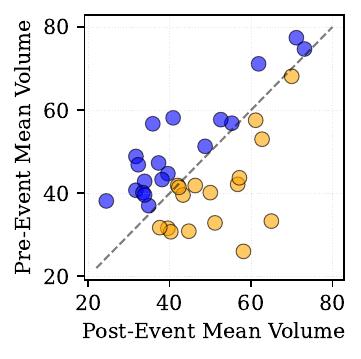}}
     \subfloat[][]{\includegraphics[width=0.23\textwidth]{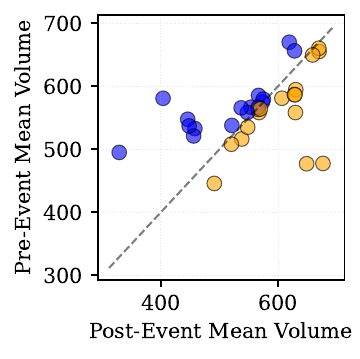}}
     
     \subfloat[][]{\includegraphics[width=.23\textwidth]{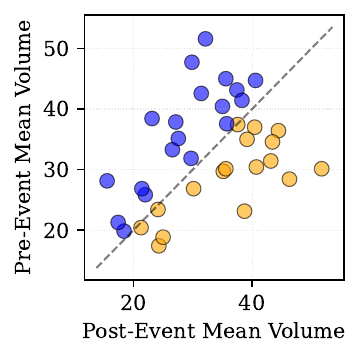}}
     \subfloat[][]{\includegraphics[width=.23\textwidth]{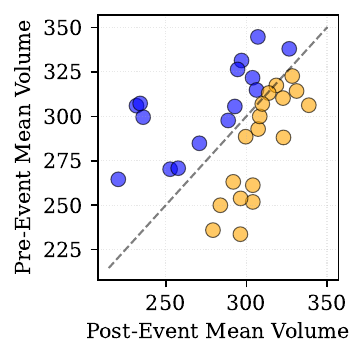}}
     
     \caption{Pre-event vs. post-event discourse volume for individual events. Each point represents one key political event (KPE), with blue indicating anticipatory patterns (pre $>$ post) and orange indicating reactive patterns (post $>$ pre). The diagonal line marks equality (pre = post).}

     \label{fig:pre_post_scatter}
\end{figure}

\section{Multi-Layer Filtering Strategy}\label{app:ml_stra}

Table~\ref{tab:data} summarizes the document counts across all layers for both Reddit posts and news articles related to \#MeToo and BLM. Some layers contain no documents for certain SM-Platform combinations: this occurs when the keyword co-occurrence patterns do not yield any documents meeting that specific threshold. For our empirical analyses in Section~\ref{sec:case_studies}, we use layer $L_5$ for both SMs and Platforms, as this layer provides a balance between semantic relevance and sufficient sample size across all datasets. The distribution of documents across layers for each SM-Platform combination is shown in Appendix (Figure~\ref{fig:source-sm-level}).

Figure~\ref{fig:source-sm-level} illustrates the cumulative distribution of documents across filtering layers for both movements and platforms. The curves reveal distinct collection patterns: \#MeToo news media shows rapid accumulation in lower layers, suggesting concentrated discourse around core keywords, while BLM Reddit exhibits more gradual accumulation, indicating broader semantic dispersion. The selection of layer $L_5$ for our empirical analyses captures sufficient semantic breadth while maintaining manageable sample sizes across all four dataset combinations.

\begin{figure}[t]
    \centering
    \includegraphics[width=0.6\linewidth]{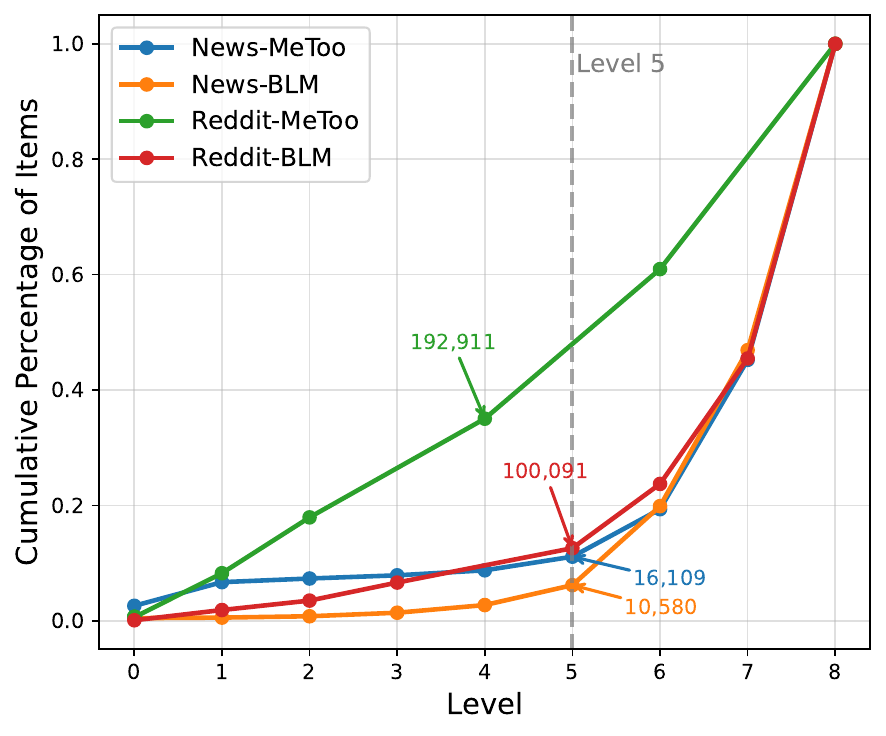}
    \caption{Cumulative distribution of documents across filtering layers. X-axis represents filtering layers ($L_0$ through $L_8$, Y-axis shows the cumulative percentage of documents. Each line represents a source-movement combination: \#MeToo News (purple), \#MeToo Reddit (pink), BLM News (teal), and BLM Reddit (cyan). Annotations indicate the number of documents at layer $L_5$, which is used for our analyses. \label{fig:source-sm-level}}
\end{figure}

\section{Event-Level Heterogeneity}
Figure~\ref{fig:pre_post_scatter} displays the relationship between pre-event and post-event discourse volume for all 36 KPEs across the four dataset combinations. Each panel reveals substantial event-level heterogeneity that is obscured in aggregate analyses.

\section{Key Political Events: Complete Listing and Categorization}
Table~\ref{tab:KEEs} provides the complete list of the 36 journalist-defined Key Political Events analyzed in this study, along with their categorization into three domains: Elections/Political Power, Foreign Policy, and Domestic Policy. Events were manually categorized by our journalist co-authors based on their primary policy focus, recognizing that some events span multiple domains.

\begin{table*}[t] 
 \caption{Key Political Events (KPEs) selected by our journalist co-authors for the case studies. Descriptions are sourced from \url{www.onthisday.com}.} 
 \label{tab:KEEs} 
 \centering 
 \footnotesize
 \begin{tabularx}{\textwidth}{r|>{\raggedright\arraybackslash}X}
 \toprule 
 \textbf{Date} & \textbf{Description} \\ 
 \midrule 
 \rowcolor{gray!10}\multicolumn{2}{c}{\textbf{Elections, Political Power \& Institutional Control}} \\ \midrule
 Sep 15, 2024 & Man arrested and charged with the attempted assassination of former US President Donald Trump, spotted while the presidential candidate was playing golf \\
 Sep 24, 2024 & US President Joe Biden addresses the UN for the last time, calling on Israel and Hamas to come to a ceasefire in Gaza, saying "I truly believe we're at another inflection point in world history" \\
 Oct 1, 2024 & US vice presidential debate between Democrat Tim Walz and Republican J.D. Vance is held in New York \\
 Oct 2, 2024 & New Justice Department indictment against President Donald Trump alleges he tries to subvert the 2020 transfer of power, taking into account Supreme Court's immunity ruling for presidents \\
 Nov 5, 2024 & Former Republican President Donald Trump is re-elected, defeating sitting Democratic Vice President Kamala Harris to become only the second president elected to non-consecutive terms after Grover Cleveland in 1884 and 1892, and the oldest elected \\
 Nov 12, 2024 & President-elect Donald Trump selects entrepreneur Elon Musk and former presidential candidate Vivek Ramaswamy, to head a newly-created Department of Government Efficiency (DOGE) \\
 Nov 14, 2024 & US Republicans re-gain control of the House of Representatives, they now control all three elected parts of US federal government - the House, the Senate, and the Presidency \\
 Nov 25, 2024 & US Justice Department special counsel Jack Smith requests and is granted dismissal of pending criminal charges against Donald Trump, based on their policy that indicting or trying a sitting president would violate the Constitution and interfere with the working of the executive branch \\
 Jan 10, 2025 & US President-elect Donald Trump is sentenced by a federal NY court to no punishment after being convicted (May 2024) of falsifying business records in a hush money case \\
 Jan 20, 2025 & Republican Donald Trump is inaugurated as the 47th President of the United States of America, and JD Vance as the 50th Vice-President; President Trump is oldest person to take the office, and joins Grover Cleveland as only the second elected to non-consecutive White House terms, the event is held indoors due to inclement weather \\
 Jan 24, 2025 & US President Donald Trump fires over a dozen independent Inspectors General, without the required 30-day advance notice citing reasons for the firing to Congress \\
 Jan 27, 2025 & James McHenry, the acting US Attorney General, fires more than a dozen prosecutors who worked for on special counsel Jack Smith's prosecution of President Trump \\
 Jan 30, 2025 & Former US Senator Bob Menendez of New Jersey receives an 11-year prison sentence for his conviction of bribery while in office \\
 Jan 31, 2025 & US Department of Justice fires more than a dozen federal prosecutors who investigated the January 6 riot, and many of the FBI agents who investigated the riot or participated in the search for classified documents at President Trump's Mar-a-Lago home \\ \midrule
 \rowcolor{gray!10}\multicolumn{2}{c}{\textbf{Foreign Policy \& International Affairs}} \\ \midrule
 Feb 18, 2025 & The United States and Russia hold talks in Riyadh, Saudi Arabia over the war in Ukraine without a representative from Ukraine \\
 Feb 20, 2025 & President Trump supporter and counterterrorism official Kash Patel is narrowly confirmed as FBI Director by the US Senate \\
 Feb 28, 2025 & Extraordinary meeting between US President Donald Trump and Ukrainian President Volodymyr Zelensky at the White House is curtailed after an argument in front of reporters \\
 Mar 3, 2025 & President Donald Trump, speaks to US Congress in the longest-ever speech by a president at 99 minutes, including a vow the US will acquire Greenland "we're going to get it - one way or the other" \\
 Apr 2, 2025 & US President Donald Trump announces "Liberation Day", unveiling wide-ranging tariffs on foreign countries importing into the US, including 34\% on China and 20\% for the European Union \\
 Apr 9, 2025 & US President Donald Trump announces reversal on global tariffs, proposing a 90-day pause–excluding China \\
 May 4, 2025 & US President Donald Trump announces 100\% tariffs on foreign films brought into the United States \\
 Aug 15, 2025 & Russian President Vladimir Putin meets with a western head of state for the first time since his country's 2022 invasion of Ukraine, hosted by US President Donald Trump at Joint Base Elmendorf-Richardson in Anchorage, Alaska; meeting to discuss framework for a peace process adjourns earlier than anticipated and no agreements are announced \\
 Aug 18, 2025 & Leaders from across Europe join Ukrainian President Volodymyr Zelenskyy at a hastily arranged summit meeting with US President Trump at the White House \\ \midrule
 \rowcolor{gray!10}\multicolumn{2}{c}{\textbf{Domestic Policy, Governance \& Social Unrest}} \\ \midrule
 Jan 4, 2025 & President Joe Biden awards the US Presidential Medal of Freedom to 19 people including chef-humanitarian José Andrés, politician Hillary Clinton, Vogue magazine editor Anna Wintour, actor Denzel Washington, soccer player Lionel Messi, actor Michael J. Fox, singer Bono, and philanthropist George Soros \\
 Feb 5, 2025 & US President Donald Trump signs an executive order banning transgender women from competing in female sports \\
 Feb 6, 2025 & US President Donald Trump reiterates that Gaza could be developed into a "Riviera of the Middle East" by the US, with its population moved elsewhere \\
 Feb 12, 2025 & Tulsi Gabbard confirmed as Director of National Intelligence by the US Senate \\
 Feb 13, 2025 & Robert F. Kennedy Jr. is confirmed as Health and Human Services secretary by the US Senate, after concerns over his anti-vaccine views \\
 Feb 21, 2025 & US President Donald Trump fires US Air Force Chief of Staff General Charles Q. Brown as Chairman of the Joint Chiefs of Staff as part of a Pentagon shake up \\
 Jun 6, 2025 & US Immigration and Customs Enforcement (ICE) sweeps in Los Angeles spark protest demonstrations \\
 Jun 7, 2025 & US President Donald Trump orders deployment of 2,000 National Guard troops to southern California to quell protests against Immigration and Customs Enforcement (ICE) sweeps in and around Los Angeles \\
 Jun 9, 2025 & US Secretary of Defense Pete Hegseth deploys 700 US Marines to protect National Guard troops in southern California who were sent to quell protests against Immigration and Customs Enforcement (ICE) sweeps in and around Los Angeles \\
 Jun 14, 2025 & Series of 'No Kings' protests occur in all 50 states across the United States, opposing the President Trump administration on the same day as the United States Army military parade in Washington, D.C. \\
 Jul 4, 2025 & US President Donald Trump signs his 'One Big Beautiful Bill' into law, narrowly passed by Senate, 51-50, and by House of Representatives, 218-214; 5 Republicans joined all Democrats in opposition \\
 Aug 12, 2025 & US President Donald Trump deploys at least 800 National Guard troops in Washington, D.C., and declares federal control of the DC police force \\
 Aug 22, 2025 & FBI raids home and office of John Bolton, former White House national security advisor to and current vocal critic of the president \\
 \bottomrule 
 \end{tabularx} 
\end{table*}

\end{document}